# PREDICTING CHANGES IN VISUAL APPEARANCE OF PERIODIC SURFACE FROM BRDF MEASUREMENTS

**Turbil C.**[1], Gozhyk I.[1], Teisseire J.[1], Simonsen I.[1], Ged G.[2], Obein G.[2]

[1] Surface du Verre et Interfaces, UMR 125 CNRS/Saint-Gobain, F-93303 Aubervilliers, France
[2] Laboratoire Commun de Métrologie LNE-Cnam, Trappes, France

colette.turbil@saint-gobain.com

**Abstract**

The present study focuses on the optical properties of functionalized surfaces and how the surface geometry impacts them. Physical measurements of reflected light are required to understand the visual aspect of such surfaces. Bidirectional Reflection Distribution Function (BRDF) is evaluated in order to identify and understand physical effects induced by different surface functionalization. BRDF measurements of high angular resolution allow us to observe diffraction phenomenon at an unusual scale. Experimental results were compared to theoretical calculation, and then diffraction effect was confirmed. This study stresses that diffractive phenomenon could have an impact on visual aspect even for simple pattern geometry. In the same way, if more complex periodic patterns are considered, as multistate periodicity pattern for example, it could rise to important modification of surface visual aspect.

*Keywords*: *Surface patterning, glossy behaviour, high resolution BRDF measurements, diffraction effect.*

**Introduction**

Carefully chosen surface modification can provide new properties for applications in wetting, haptics, optics to almost any surface [1]. Yet, surface modification can also impact visual properties and thus aesthetics of the product. Visual appearance is a major criterion for customers: the final aspect of surface could be an issue for surface patterning. Link between surface pattern, physical measurement of light reflection such as BRDF (Bidirectional Reflection Distribution Function) and visual appearance is not trivial [2]. The goal of our study is to bring first pieces of knowledge in the comprehension of the relation between surface patterning, BRDF and visual appearance of such functionalized surfaces. In order to understand all phenomena at stake, high angular resolution BRDF measurements are required to be compared to theoretical calculations based on grating formula.

**1  Samples description**

In the sake of simplicity, optical properties of hydrophobic surfaces are studied on the example of model hydrophobic patterns with simple geometry (Fig.1). Each sample consists of a glass slide with the patterned hybrid silica layer on its top (Fig.1a). Used patterns were cylindrical micro pillars (height (*h*) and diameter (*d*): 10 µm) organized within hexagon lattice. The lattice constant (*a*) varied from 20 to 80 µm (Fig.1b), which in terms of the *surface coverage rate*

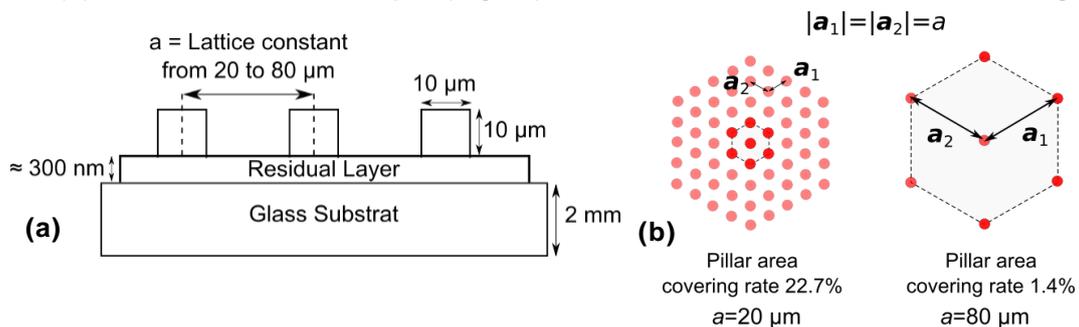

**Figure 1 – Schematic description of the geometry of samples: a) stack of layers (side view), b) top view on hexagonal gratings of different lattice constant (20 and 80 µm)**

$\rho$ (surface parameter defining the wetting properties [3]) represents the variation within the range between 1.4% (80 µm lattice) and 22.7% (20 µm lattice). In case of hexagonal lattice this latter parameter is calculated by the following formula:

$$\rho = \pi \frac{d^2}{a^2} \frac{\sqrt{3}}{6} \tag{1}$$

Residual sol-gel layer between the glass surface and pillars is necessary to provide the adhesion of pillars to the surface. 300 nm thickness of this layer was chosen to prevent from optical guiding effects.

Device fabrication process consists of two main steps: silica sol-gel chemistry and nanoimprint (for detailed description see [4]). A hybrid silica sol-gel is prepared and is spin coated on a glass substrate (5 cm x 5 cm in our case). Then the nanoimprint step is applied: a PDMS (Polydimethylsiloxane) mold, previously patterned, is put on the coated glass substrate. The pattern is then replicated from the mold to the sol-gel layer under controlled conditions (1 bar pressure and temperature of 110°C during 45 minutes). This fabrication process allows creating various types of functionalized surfaces on glass substrate, with high reproducibility.

All samples were examined in a "Spectralight QC X-Rite" light booth, under D65 illumination. Surface patterning only slightly modifies transmission properties of glass-based samples. Nevertheless there are significant visual differences in reflection between samples with relatively high (>8%, *a*=20 and 30 µm) and lower surface coverage (≤5%, *a*≥40µm) and especially at grazing angles (Fig.2). Meanwhile, samples with higher coverage rate (>8%, *a*=20 and 30 µm) look diffuse-like in reflection, as they exhibit a white-coloration typical for lambertian-like rough surfaces. Samples with low surface coverage (≤5%, Fig.2a) give the impression of smooth and specular-like surfaces (as illustrated on Fig.2b, there is no blurring on the image of coloured grid reflected by such surfaces). These observations suggest that surface coverage rate of such patterned surfaces impact their gloss.

In order to understand visual rendering of samples, optical properties have to be examined. Integral entities such as gloss and haze are not sufficient and spatially resolved measurements should be applied to understand this effect. Due to the highly glossy behaviour of samples (especially for low surface coverage), a goniospectrophotometer with high angular resolution is required to perform relevant measurements.

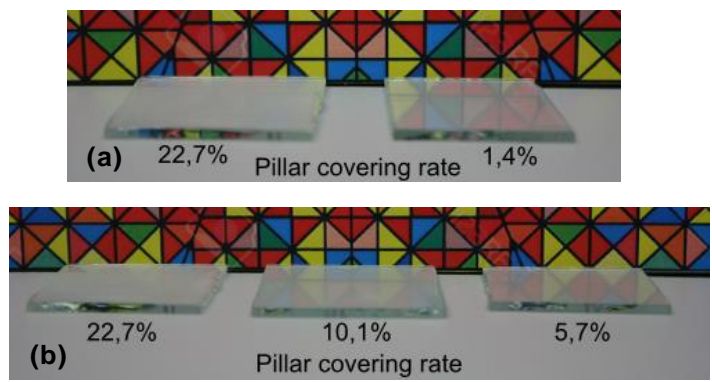

**Figure 2 – Photographs of examined samples taken in a light booth (Spectralight QC X-Rite) under D65 illumination at grazing angles: a) comparison of samples with the highest (22.7%) and the lowest (1.4%) available surface covering rates; b) comparison of sample with the highest surface covering rate with samples with intermediate surface covering values.**

## 2 Theory of grating formula

As light scattering phenomena is typically represented as a function of polar and azimuthal angles, an angular convention, used to describe illumination and detection angles on this paper, is presented on Fig.3, adapted from [5]. $\theta_i$ and $\phi_i$ describe illumination direction while $\theta_r$ and $\phi_r$ describe reflected direction.

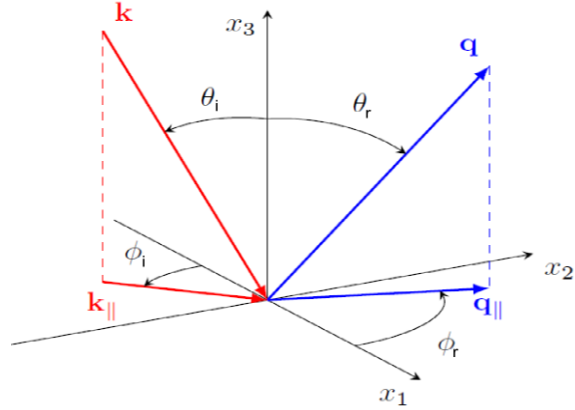

**Figure 3 – Geometrical description of illumination and reflected angles: $\theta_i$ and $\phi_i$ describe illumination direction and $\theta_r$ and $\phi_r$ describe detection direction.**

We now turn to the light scattered from the samples introduced in the previous section. The phenomenon of light scattering from periodic surface structures is known as diffraction, and it has been studied for centuries, and it is still of contemporary interest in science and engineering [7]. When incident light is scattered from a periodic structure, known as a grating, the reflected light will typically consist of a set of pronounced peaks in the reflected intensity distribution. Each of these peaks is referred to as diffractive orders, and the positions of these peaks are mainly determined by the wavelength of the incident light, the type of grating and parameters defining it.

Mathematically these angular positions are determined by the grating formula [8] that we now will introduce (in a pedagogical fashion). To this end, we will assume the scattering geometry depicted in Fig. 3. The planar surface supporting the pillars is assumed to coincide with the $x_1 x_2$ plane. In the region $x_3 < 0$ we have a medium that is characterized by the dielectric constant $\varepsilon$, that is assumed to be real and positive (i.e. a dielectric); and in the region that is $x_3 > 0$ and outside the pillars, we assume to have vacuum ($\varepsilon_0 = 1$). Into this structure, light of wavelength $\lambda$ is incident at the angles of incidence ($\theta_i$, $\phi_i$). The component of the incident wave vector $\boldsymbol{k}$ that is parallel to the planar surface of the substrate is then

$$\boldsymbol{k}_{\|} = \frac{\omega}{c} \sin\theta_i (\cos\phi_i, \sin\phi_i, 0), \tag{2a}$$

where we have introduced the frequency $\omega$ of the incident light, and $c$ denotes the speed of light in vacuum [$\omega/c = 2\pi/\lambda$]. Similarly, the lateral wave vectors of a beam reflected into the angles of reflection ($\theta_r$, $\phi_r$) are (see Fig. 3)

$$\boldsymbol{q}_{\|} = \frac{\omega}{c} \sin\theta_r (\cos\phi_r, \sin\phi_r, 0), \tag{2b}$$

The lattice of the periodic structure is invariant under a translation by the vector $\boldsymbol{x}_{\|}^{(l)} = l_1 \boldsymbol{a}_1 + l_2 \boldsymbol{a}_2$, where $\boldsymbol{a}_i$ (i = 1; 2) are the two non collinear primitive translation vectors (in the $x_1 x_2$ plane) that characterize the lattice (see Fig. 1), $\boldsymbol{l} = (l_1, l_2)$ and $l_i$ is an integer ($l_i \in \mathbb{N}$). The area of the unit cell is $A_c = |\boldsymbol{a}_1 \times \boldsymbol{a}_2|$. Associated with the direct lattice, characterized by $\boldsymbol{a}_1$ and $\boldsymbol{a}_2$, is the reciprocal lattice defined by the translation vectors

$$\boldsymbol{G}_{\|}^{(m)} = m_1 \boldsymbol{b}_1 + m_2 \boldsymbol{b}_2, \tag{3}$$

where **m**=(m₁,m₂), mᵢ ∈ ℕ, and the primitive translation vectors, **b**ᵢ, are defined from the relation (i, j = 1, 2)

$$\boldsymbol{a}_i \cdot \boldsymbol{b}_i = 2\pi\delta_{ij}, \tag{4}$$

where $\delta_{ij}$ represents the Kronecker delta.

In this work we assume that the direct lattice formed by the pillars is of the hexagonal type, so that the reciprocal lattice also is hexagonal. Without loss of generality, we orient the coordinate system so that the primitive lattice vectors may be taken as

$$\boldsymbol{a}_1 = a\frac{\sqrt{3}}{2}\hat{\boldsymbol{x}}_1 + a\frac{1}{2}\hat{\boldsymbol{x}}_2 \qquad \boldsymbol{a}_2 = -a\frac{\sqrt{3}}{2}\hat{\boldsymbol{x}}_1 + a\frac{1}{2}\hat{\boldsymbol{x}}_2, \tag{5}$$

where *a* > 0 is the *lattice constant* and a caret over a vector indicates a unit vector. The corresponding primitive translation vectors for the reciprocal lattice, *bᵢ*, are determined from Eq.(4).

After the incident light, characterized by the wave vector $\boldsymbol{k}_\parallel$ [Eq. (2a)], has interacted with the periodic structure, the diffracted wave of order **m** in either reflection or transmission will have the wave vectors which component is parallel to the surface of the substrate

$$\boldsymbol{q}_\parallel^{(\boldsymbol{m})} = \boldsymbol{k}_\parallel + \boldsymbol{G}_\parallel^{(\boldsymbol{m})}. \tag{6}$$

Equation (6) is the grating formula that we seek, and it works equally well in both reflection and transmission. It should be noticed that for a given diffractive order **m**, the wave vectors $\boldsymbol{q}_\parallel^{(\boldsymbol{m})}$ does not necessarily describe a propagating mode (an "open channel"); only those modes in reflection that satisfies $q_\parallel^{(m)} = \left|\boldsymbol{q}_\parallel^{(m)}\right| < \omega/c$ are propagating (in the ambient medium), and only when $q_\parallel^{(m)} = \left|\boldsymbol{q}_\parallel^{(m)}\right| < \sqrt{\varepsilon}\omega/c$ is a mode propagating in transmission (assuming ε positive).

To relate an open diffractive channel in reflection to the angles of reflection (θr(**m**), φr(**m**)) that defines it, we have to equate the right-hand sides of Eq. (2b) and Eq. (5) $\boldsymbol{q}_\parallel = \boldsymbol{q}_\parallel^{(m)}$ and solve the resulting equation system for those angles. The results are

$$\theta_r^{(\boldsymbol{m})} = arcsin\left(\frac{\left|q_\parallel^{(m)}\right|}{\omega/c}\right), \qquad \left|\boldsymbol{q}_\parallel^{(m)}\right| < \frac{\omega}{c}, \tag{7a}$$

and

$$\phi_r^{(\boldsymbol{m})} = \arctan(q_2^{(m)}, q_1^{(m)}), \qquad -\pi < \phi_r^{(m)} \leq +\pi. \tag{7b}$$

In obtaining Eq. (7) we have used that $q_\parallel = (\omega/c)sin\theta_r$, and the two argument arctangent function, denoted arctan(y,x) is used in order to determine the appropriate quadrant of the resulting angle (and to properly treat the case when $q_1^{(m)} = 0$). In most computer languages this function is named atan2(.,.).

## 3 Materials and methods

BRDF measurements were realized with ConDOR (Conoscopic Device for Optical Reflectometry), a goniospectrophotometer with very high angular resolution (0.018°) dedicated to BRDF measurements within the specular zone. Measurements were performed using *spatial multiexposure and time multiexposure*. Spatial multiexposure allows to enlarge the range of detection of ConDOR setup from 2° (±1° around the specular direction) to 8° (±4°). Time multiexposure method provides high dynamics range (due to dealing with saturation issues of the CCD detector) and thus enables to detect the entire specular peak. Complete details on ConDOR goniospectrophometer as well as spatial and time multiexposure techniques are published elsewhere [6].

Samples with following covering rates were measured: 22.7%, 10.1%, 3.6% and 1.4%. Angular configurations of illumination and detection were fixed to: $\theta_i=\theta_r=45°, \phi_i=\phi_r=0°$. A V($\lambda$) filter was used during the measurements in order to reproduce the spectral sensitivity of human visual perception.

Gratings formula, described in the previous section, was implemented in order to predict the angular position of the diffractive orders. Illumination parameters were fixed to $\theta_i=45°$ and $\phi_i=0°$.

## 4 Results

### 4.1 BRDF measurements with ConDOR

All BRDF measurements exhibit a common structure of the spatial distribution of reflected light as shown in Fig.4: a central specular peak surrounded by several secondary peaks, organized according to a hexagonal lattice. Angular distance between specular peak and secondary peaks varies as a function of the pattern's size: a large period pattern induces a small angular distance, as presented in Fig.5. Plane cuts presented in Fig.5 were realized by representing the plane (red line) defined by the position of the specular peak and of one secondary peak (yellow crosses). For $a$=20 µm and $a$=80 µm, angular distances between the specular peak and its nearest neighbor secondary peak are respectively, 2.1° and 0.56°.

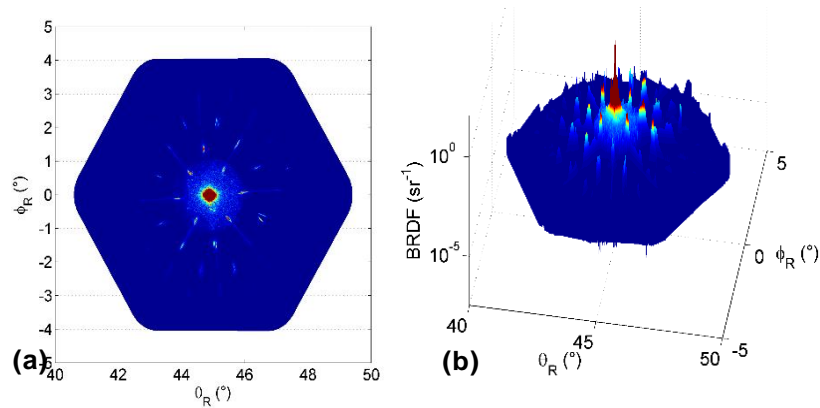

**Figure 4 – BRDF measurement of sample a=50 µm at logarithmic scale: a) top view; b) a 3D representation. Experiment performed at $\theta_i=\theta_r=45°, \varphi_i=\varphi_r=0°$ under unpolarized illumination with V($\lambda$) filter.**

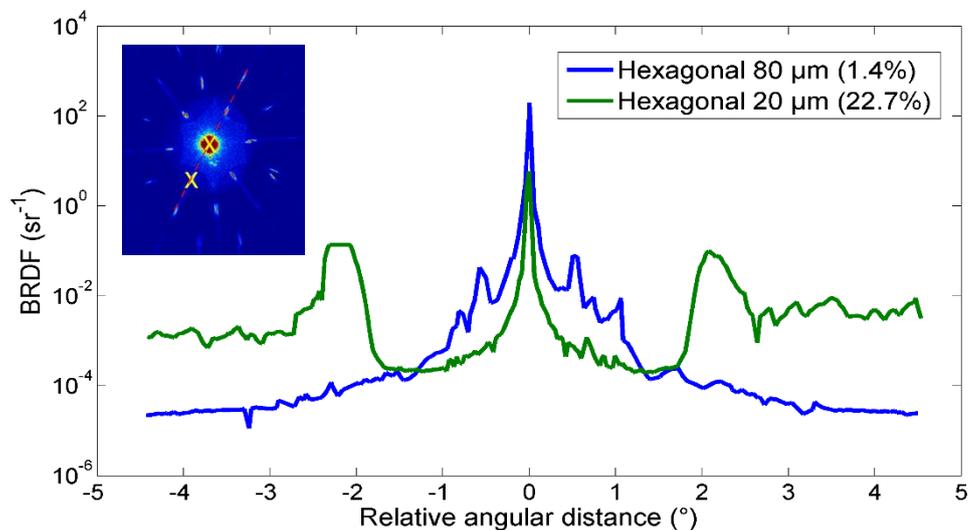

**Figure 5 – Comparison of BRDF data from samples with different lattice constants (a=20 and 80 µm). Plane cuts are given in logarithmic scale.**

The maximum of the specular peaks for all BRDF measurements are reported on Fig.6a. BRDF values vary from 5.91 sr$^{-1}$ (for the densest pattern) to 241.6 sr$^{-1}$ (the least dense pattern). As expected low surface covering rate leads to high level of BRDF. BRDF values of secondary peaks are also influenced by pillar covering rate. Fig.6b represents the ratio between BRDF values of specular peak and the averaged secondary peaks as a function of covering rate. The increase in covering rate induces the decay in this ratio, meaning that secondary peaks become more and more intense (this can be also observed on Fig.5). Pattern with small covering rate reflects light close to the specular direction (in a +/- 1° window), and the intensity of the specular peak makes the major contribution to BRDF compared to secondary peaks. While in case of the pattern with high covering rate the light is diffused in a wider angular range and the intensity of secondary peaks increases significantly.

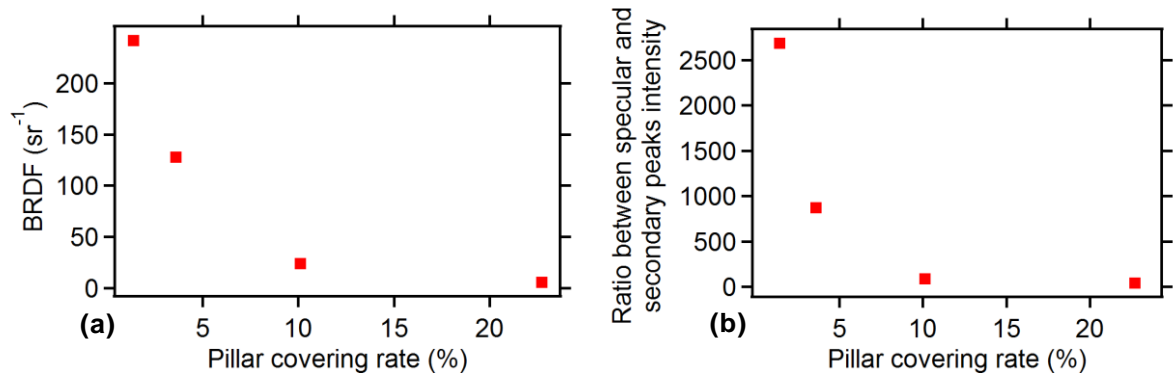

**Figure 6 – BRDF values as function of pillar covering rate: a) maximum of the specular peak, b) ratio between BRDF peak values in specular direction and averaged peak value of secondary peaks.**

### 4.2 Theoretical calculation results and comparison to measurements

As mentioned in Section 2, spatial distribution of the light reflected from the samples contains different diffractive orders defined by two angles (polar and azimuthal angles). Those angular positions are shown in Fig.7a and Fig.7b, respectively for patterns with *a*=20 µm and 80 µm. As expected, those calculations predict a hexagonal organization of diffractive orders, and a strong relation between the pattern period and the angular distribution of diffractive orders. For instance, angular distance between two neighbors diffractive orders for *a*=20 µm pattern and 80 µm pattern are respectively 2.16° and 0.544°.

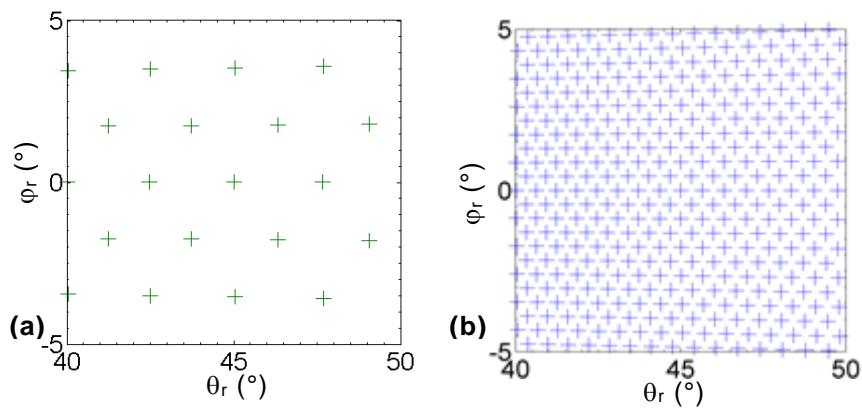

**Figure 7 – Calculated angular positions of diffractive orders for a=20 µm (a) 80 µm (b). Calculation performed with parameters of the experiment: $\theta_i$=45°, $\varphi_i$=0° and $\lambda$=555 nm**

Measurements and calculations are compared on the example of the sample with 3.6% covering rate (Fig.8). Fig.8a shows a good correlation between measurements (yellow dots) and calculations (blue crosses) and confirms the presence of diffraction phenomenon. However, a small misalignment is observed resulting in small angular shift between experimental data and calculations. It is corrected in Fig.8b through the 5° turn in azimuthal angle of illumination ($\phi_i=5°$). We attribute it to the inevitable misalignment of the micrometric pattern of the sample with the respect to the incidence plane. We finally end up with a very good angular correlation presented on Fig.8b.

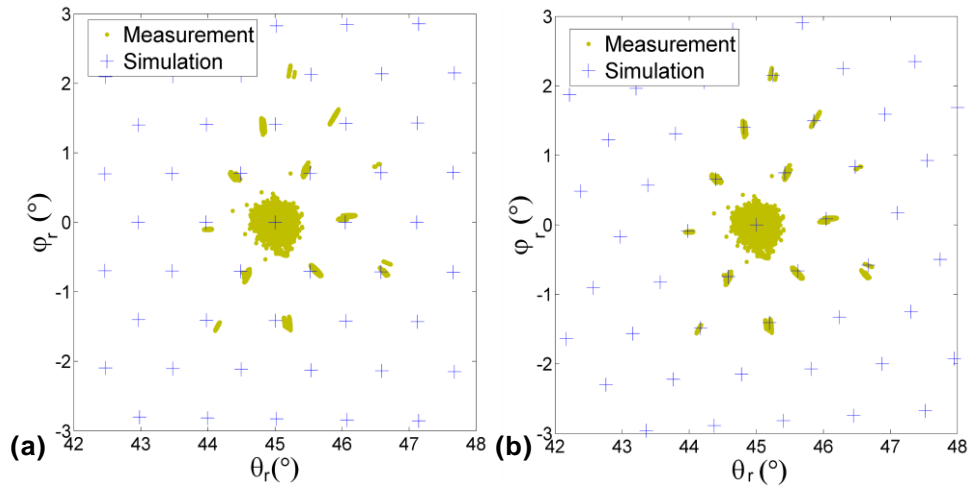

**Figure 8 – Positions of calculated (blue crosses) and experimentally observed (yellow points) diffractive orders: a) Illumination configuration are $\theta_i=45°$ and $\phi_i=0$ for both measurements and calculations; b) calculation at $\theta_i=45°$ and $\phi_i=5°$ (to reproduce the misalignment of the pattern).**

## 5  Discussion

Large gratings ($a$=80 μm in our case) induce an intense specular peak (241 sr$^{-1}$), surrounded by several secondary diffractive peaks with a weaker intensity (2500 times smaller). All those peaks are localized in a ±0.75° window around the specular direction. As it is schematically shown in Fig.9a, such patterns should enhance gloss property, due to their highly localized response in the specular direction. Dense gratings ($a$=20 μm for this study) show a different behaviour. A specular peak (5.9 sr$^{-1}$) is surrounded by only 40 times smaller secondary peaks which appear within a larger (±4°) window. With the same schematic representation, shown in Fig.9b, the gloss zone is both enlarged and attenuated. Thereby, such diffractive pattern enhances the non-specular reflection of light resulting in the enhanced haze property. Finally, the variation of angular separation between specular peak and secondary peaks (as well as the variation in the relative intensity of diffractive orders) with the lattice period raise questions on the definition of gloss zone.

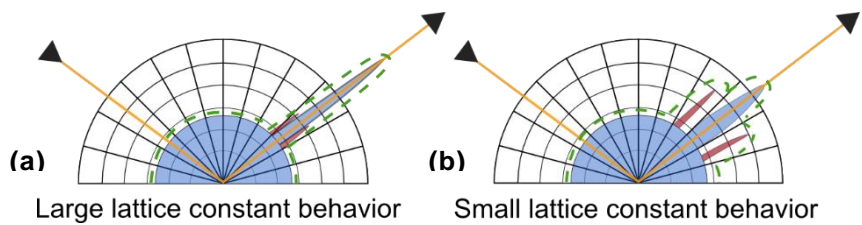

Large lattice constant behavior    Small lattice constant behavior

**Figure 9 – Schematic representation of the gloss behaviour of patterned samples with different periods: a) large period and b) small period.**

The diffraction phenomenon is rarely considered in the analysis of visual appearance. However, the experimental data presented in this paper prove that diffraction effect should not be neglected in case of surface functionalization with periodic patterns (rather common for nano-imprint patterning technique). Moreover, patterns described here present a simple organization and the more complex lattice shape or multi-scale patterns could induce a stronger optical response and may significantly modify surface visual aspect. Furthermore, as rendering models based on microfacets theory do not account for light diffraction, they will fail to produce realistic visual rendering at least for the surfaces with periodic patterns.

# 6 Conclusion

Functionalization through surface patterning may significantly influence the visual properties of the object, especially in case of transparent objects. In the case of surface patterning with periodic arrays of micropillars, spatially resolved measurements showed the presence of numerous diffraction orders with angular separation and relative intensity stretching with the increase of surface density of pillars. Diffractive effects can induce slight modifications of visual aspect of such surfaces. Large pattern period tends to enhance gloss property of surface due to the localized angular distribution of diffractive peaks in the specular direction; while pattern with smaller period induces a larger angular distribution leading to a more diffuse surface aspect. Furthermore, those effects could be dramatically enhanced when dealing with more complex lattice patterns, consequently diffraction phenomenon must not be neglected when visual aspect of periodic system is studied.

## Acknowledgement

Authors express deep gratitude to A.M. Rabal (LCM, LNE/CNAM, Trappes, France) for her precious help with ConDOR measurements.